\documentclass[journal]{IEEEtran}
\usepackage[utf8]{inputenc}

\ifCLASSINFOpdf
  \usepackage[pdftex]{graphicx}
  \DeclareGraphicsExtensions{.pdf,.jpeg,.png}
\else
  \usepackage[dvips]{graphicx}
  \DeclareGraphicsExtensions{.eps}
\fi

\usepackage{listings}

\lstdefinelanguage{json}
{
  morestring=[b]",
  morestring=[d]'
}

\usepackage{array}

\ifCLASSOPTIONcompsoc
  \usepackage[caption=false,font=normalsize,labelfont=sf,textfont=sf]{subfig}
\else
  \usepackage[caption=false,font=footnotesize]{subfig}
\fi

\usepackage{stfloats}
\usepackage{color}
\newcommand{\revision}[1]{{#1}}

\usepackage{url}

\begin{document}
%
\title{Automatic Intent-Based Secure Service Creation through a Multilayer SDN Network Orchestration}
%
%
%

\author{Thomas Szyrkowiec, Michele Santuari, Mohit Chamania, Domenico Siracusa, Achim Autenrieth, Victor Lopez, Joo Cho and Wolfgang Kellerer%
\thanks{T. Szyrkowiec, M. Chamania, A. Autenrieth and J. Cho are with ADVA Optical Networking SE, Germany e-mail: tszyrkowiec@advaoptical.com}%
\thanks{M. Santuari and D. Siracusa are with CREATE-NET Research Center, Italy}%
\thanks{V. Lopez is with Telefonica I+D, Spain}%
\thanks{W. Kellerer is with the Chair of Communication Networks at Technical University of Munich, Germany}%
\thanks{\newline \copyright 2018 Optical Society of America. Users may use, reuse, and build upon the article, or use the article for text or data mining, so long as such uses are for non-commercial purposes and appropriate attribution is maintained. All other rights are reserved.}
}

\maketitle

\begin{abstract}
Growing traffic demands and increasing security awareness is driving the need for secure services.
Current solutions require manual configuration and deployment based on the customer's requirements.
In this work, we present an architecture for an automatic intent-based provisioning of a secure service in a multilayer --- IP, Ethernet and optical --- network while choosing the appropriate encryption layer using an open-source SDN orchestrator.
The approach is experimentally evaluated in a testbed with commercial equipment.
Results indicate that the processing impact of secure channel creation on a controller is negligible.
As the time for setting up services over WDM is varying between technologies, it needs to be taken into account in the decision process.
\end{abstract}


%
\IEEEpeerreviewmaketitle

\section{Introduction}
%
%
%
%
\IEEEPARstart{I}{nternet} proliferation has increased exponentially in the last two decades, and it is estimated that 40\% of the population or more than 3.5~billion people have Internet access \cite{livestats}.
The ability to reach a significant global population base has been the primary driver for businesses to provide essential services over this infrastructure.
However, companies have to contend with higher risk and potential costs associated with data breaches.
A recent study estimates the average potential cost of a data breach to be as high as \$3.6~million \cite{databreach}.
As a result, it is critical to deploy solutions to secure the distributed cyber infrastructure.
Network encryption is a key component in the cyber-security environment and is responsible for ensuring that communication between two trusted endpoints cannot be intercepted by malicious attackers.
Network encryption is crucial for communication mechanisms operating over an untrusted public infrastructure.
Consequently, protocols such as Hypertext Transfer Protocol Secure (HTTPS) and Secure File Transfer Protocol (SFTP) natively support encryption.
However, as applications move from dedicated physical infrastructure to distributed and virtualized infrastructure in the cloud, many communication protocols, that do not natively support encryption, can potentially be exploited by malicious attacks.

Given the large number of communication protocols, a deployment of specialized mechanisms for each individual protocol is not feasible and in-flight encryption is used as a standard mechanism to secure these protocols.
In-flight encryption is applied to traffic on one of the lower layers of the OSI model, i.e., physical (L1), data link (L2) and network layer (L3).
Protocol solutions operating at those network layers (e.g., IPsec\cite{IPsec}, MACsec\cite{MACsec}, physical layer\cite{ADVA}) have inherent technical (e.g., latency, effective throughput) and cost trade-offs.
In-flight encryption assumes that protocols that do not support security mechanisms will be encapsulated into one of these protocols.
Specific implementations also differ in the choice of mechanisms used for authentication, secure key exchange, payload encryption and strategies for storing encryption settings on end-devices.
All of them determine the complexity associated with breaking the encryption mechanism.

Network service providers typically deploy infrastructure with multiple, potentially vendor-specific, choices for in-flight encryption.
Manually evaluating the technical\,/\,cost and security trade-offs between the possible solutions for an application, requesting a ``secure'' connectivity service from the infrastructure, poses a significant overhead.
The evaluation process also needs to be revised as the infrastructure deployed by the service provider is changed.
Furthermore, applications are interested in satisfying their own requirements, which commonly do not match the priorities of the network operator: while applications are concerned about bandwidth, latency, availability, security, etc., management systems are optimized to minimize the usage of resources, the energy consumption and so on.
In theory, applications should be able to clearly define their needs without getting into details on how the service is created.
This corresponds to the concept of an intent.
On the other side, network control mechanisms should be designed in order to manage the different layers of the infrastructure in such a way that the parameters of interest for operators are optimized, while providing the necessary set of resources to satisfy applications’ requirements.
With respect to the scenario presented in this paper, this may imply that the type of in-flight encryption chosen for the secure service may vary according to the specific needs of the requesting application, e.g., for very low-latency applications physical layer encryption may be preferred over higher network layers.
Throughout this paper we will use the term in-flight encryption for describing the process of applying a technology to encrypt traffic.
The secure service is used in the context of a connection that is transferring traffic that needs to be secured by some kind of (in-flight) encryption mechanism.

\begin{figure*}[!t]
  \centering
  \subfloat[Encryption at the IP layer reusing existing (optical) connections between the routers.]{
    \includegraphics[width=2.5in]{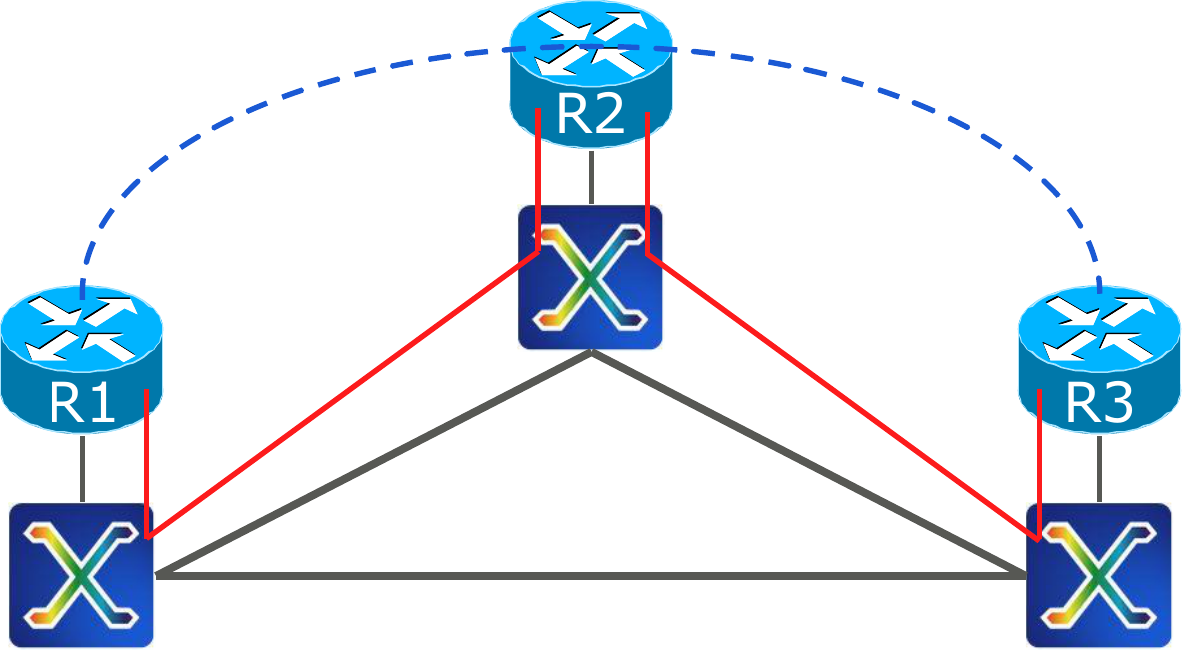}%
    \label{fig:layer_ip}}
  \hfil
  \subfloat[Encryption at the physical layer requiring the setup of a new lightpath (blue solid line).]{
    \includegraphics[width=2.5in]{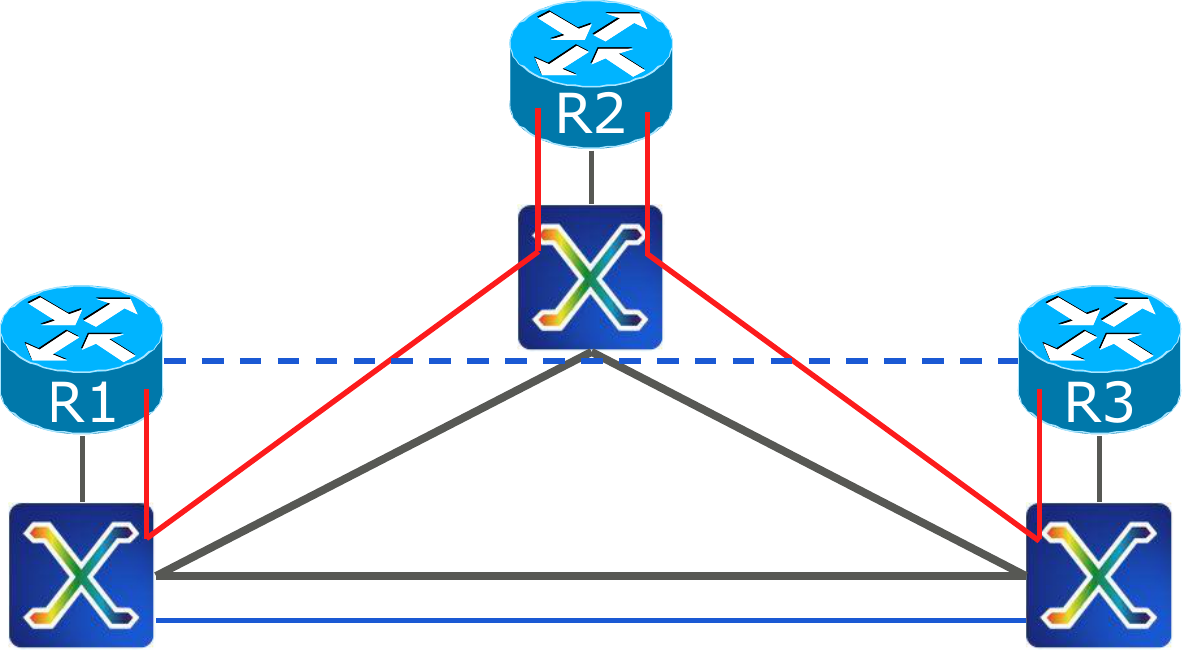}%
    \label{fig:layer_phys}}
  \caption{A network consisting of routers and optical equipment. The red lines indicate established connections between the routers. The dashed blue line represents the resulting secure service between R1 and R3 by applications with different requirements.}
  \label{fig:layer}
\end{figure*}

Applications need a way of describing requirements directed at an underlying network in a technology-agnostic manner.
The main contribution of this work is the concept of an automatic selection of the appropriate encryption layer in a multilayer IP\,/\,optical network based on requirements that are expressed by an application through intents.
The goal is to move the decision complexity away from the application requesting the service toward the orchestrator.
Intents define the application's requirements (e.g., bandwidth), cost constraints, and application type, which in turn may constrain the choice of the technology used for the secure service.
This concept is experimentally validated with an implementation using an open-source controller and commercial hardware.
The controller is responsible for receiving and translating application's intents into network requirements, evaluating the related trade-offs and constraints, and eventually provisioning a secure service that can be used by the application.
Finally, the implementation is complemented by measurements and an evaluation in a real testbed equipped with optical and Ethernet equipment.

\section{Secure Services}

Service providers used to deploy separate networks to host enterprise and end customer services.
However, such an approach is not scalable in economic terms as the infrastructure is duplicated and underutilized.
Current deployments are based on unified networks implementing an IP architecture, which means that all traffic flows share the same infrastructure.
As a result, more and more businesses are investing in improving their security infrastructure.

End-to-end encryption was used widely to protect the transfer of very sensitive information, e.g., online banking.
However, the general public has developed concerns about confidentiality on the Internet in recent years, due to e.g., open WiFis and pervasive monitoring.
This has led to a widespread utilization of HTTPS for the Internet traffic, mainly driven by the over-the-top providers.
The percentage of HTTPS and HTTP requests was 14\% - 86\%  in January 2015, 24\% - 76\% in January 2016 and 50\% - 50\% in August 2017 \cite{archive}.
End-to-end encryption is flexible and independent of the underlying network infrastructure, making it relatively easy to deploy.
This flexibility comes at the cost of higher processing requirements on both --- server and client --- communication endpoints, which consequently induces more latency and reduces the throughput of the network.
When performing such tasks at the network level, requirements on endpoint capabilities and processing complexity are relaxed.
Moreover, this allows the network operators to optimize the cost per bit for encrypting traffic between two remote sites.
Different mechanisms such as IPsec, MACsec, and custom all-optical encryption differ in the cost of deployment, availability, latency and throughput.
The selection of the best encryption mechanism for the applications is a feature that the network must provide in order to optimize the application's experience, while providing the most cost-effective solutions.
This is called a Secure Transmission as a Service \cite{EUCNC}, which can be seen as being part of the Security as a Service model \cite{secaas}.

Let us explain secure services and their assignment to network layers using an illustrative example (Fig.~\ref{fig:layer}).
We assume a regular carrier network composed of IP and optical equipment.
For the sake of simplicity, one router is collocated with every optical device and both are connected to each other.
The red lines represent an existing connectivity between the routers over optical.
The blue dashed line shows the resulting connection between the endpoints.
Finally, the blue solid line indicates a newly set up optical connection.
We consider two applications requiring a secure service between the two routers R1 and R3.
One of them has only a small bandwidth demand and does not provide any further constraints.
The secure service will be routed over an end-to-end IPsec tunnel, reusing existing optical connections (Fig.~\ref{fig:layer_ip}).
However, a second application requests not only security, but also a high bandwidth, so it will be more cost effective to use an encrypted optical circuit to minimize the network resource utilization (Fig.~\ref{fig:layer_phys}).
In general, this encrypted optical connection needs to be setup exclusively on demand.

\revision{
  Current research on security in SDN is focusing on controllers and the communication between the control and data plane \cite{sdn_sec_survey, cost_of_security, controller_security}.
  Even though the data plane is also considered in isolation, the work is mostly limited to OpenFlow switches \cite{sdn_threats}.
  Therefore, it is concerned with preserving the integrity of flow tables and flow rules \cite{sdn_sec_survey}.
  Eavesdropping is a general threat in computer networks and it applies to the electrical \cite{eavesdropping} as well as the optical domain \cite{optical_security}.
  A natural countermeasure is encryption.
  In OpenFlow networks encryption of packets may result in matching issues because some headers are no longer accessible.
  Our work tries to fill this gap by providing a solution for eavesdropping at the data plane for secure services in a multilayer environment.
  This paper is based on previous work carried out by the authors.
  In \cite{ecoc_pdp} we provided a proof of concept (PoC) based on ONOS (\url{http://onosproject.org/}) for an automatic secure service instantiation at the IP and optical layer and evaluated the overhead introduced by the processing in the controller.
  We extended this PoC by including MAC layer encryption and using a high-level intent interface developed in ACINO (\url{http://www.acino.eu/}) \cite{ofc_demo}.
  The work at hand demonstrates how ONOS' existing intent interface with minimally invasive extensions can be used to select the proper layer for encryption.
  Newly developed components for ONOS, like domain intents and a driver for transport networks, are integrated and an updated intent compiler is introduced.
  Finally, the evaluation of the compilation process and of the setup and tear down times provides insights into additional constraints for the proper selection of the encryption layer.
}

\section{Encryption mechanisms}

\begin{table*}[!t]
    \caption{Overview of encryption layer properties.}
    \label{table:layers}
    \centering
    \begin{tabular}{lccc}
        Requirement & IPsec (L3) & MACsec (L2) & Physical (L1) \\ 
        \hline
        Latency & high & medium & low \\ 
        Data Throughput & low & medium & line speed (no overhead) \\ 
        Protocol Transparency & low & medium & high \\ 
        Flexible Encrypted Payload Size & restricted & restricted (standard MAC) & 1G -- 100G \\ 
        End-to-End Compatibility & IP only & layer 2 only & Fiber\,/\,OTN or SONET\,/\,SDH \\ 
        Flexibility (Meshed) & high & low & medium \\ 
    \end{tabular}
\end{table*}

The main purpose of an encryption mechanism is to protect the user data against eavesdropping.
The choice of the encryption mechanism depends on the requirements of the application.
In this section, we describe the general principles behind in-flight encryption services, and compare various properties of physical layer encryption, MACsec and IPsec.
They are summarized in Tab.~\ref{table:layers}
These properties are used by the orchestrator to assign an appropriate encryption mechanism to the requested secure service.
The configuration of an encrypted connection is a multi-step process.
First the two endpoints need to be authenticated, which means identifying the other side as the expected communication partner.
This can be done by a pre-shared key, username\,/\,password based authentication, or using certificates.
In our evaluation, we use pre-shared keys for all mechanisms.
\revision{
  The encryption mechanism needs two cryptographic primitives: a symmetric-key algorithm and a key exchange protocol.
  A symmetric-key algorithm is used to provide data confidentiality.
  It is called ``symmetric'' because the same key is used for encryption and decryption.
  The most popular symmetric-key algorithm today is the Advanced Encryption Standard (AES), which we use in our experiments.
  A key exchange protocol, on the other hand, is based on public-key cryptography and it is used to derive a symmetric session key for the AES encrypted communication.
  The Diffie-Hellman (DH) key exchange protocol can establish a symmetric key securely over a public channel.
  The evaluated mechanisms use various incarnations of the DH protocol.
  Note that the symmetric key is regularly refreshed in order to limit the amount of data that is encrypted with the same key.
}

Physical layer encryption \cite{ADVA} is a point-to-point Layer 1 encryption over the optical fiber.
In this mechanism, the hardware encrypts the Optical Channel Data Unit (ODU) payload before inserting it into the Optical Channel Transport Unit (OTU) frame, which is decrypted at the peer site.
Because the payload of an Optical Transport Network (OTN) frame is encrypted with no additional overhead, it offers a high protocol transparency.
This approach provides the lowest latency and highest throughput, i.e., line speed, among the presented mechanisms.
The bulk data encryption is typically done by a symmetric-key crypto algorithm, such as AES, and the session key can be established by a key exchange protocol, such as authenticated DH key exchange.
The drawback of physical layer encryption is that it is bound to custom hardware at both endpoints of the infrastructure and the setup time roughly corresponds to a lightpath provisioning.

The Media Access Control Security (MACsec) is a security protocol for Ethernet links (Layer 2) and enables secure communication between neighboring nodes.
Each packet is encrypted using symmetric key cryptography so that the communication cannot be monitored or altered while traversing the link.
The symmetric key is established by a higher-level protocol by which the endpoint authentication and the key agreement are performed.
The protocol is part of IEEE Std 802.1X-2010 \cite{MACsec:auth}.
The MACsec data frame, defined by IEEE Std 802.1AE \cite{MACsec}, adds a security tag and an integrity check value to the Ethernet frame.
Both are checked by the receiver to ensure that the data has not been compromised while being transmitted over the link.

Since MACsec is a simple protocol, it is used to achieve a high speed transmission with low latency\,/\,overhead on Ethernet links.
However, when data traverses multiple hops, where MACsec encryption is enabled on the Ethernet link, each L2 device must encrypt\,/\,decrypt the frame.
This may cause performance degradation, compatibility problems and security issues, if an intermediate device is untrusted.
Also, addressing\,/\,reachability limitations of Layer 2 apply.

Internet Protocol Security (IPsec) \cite{IPsec} is an end-to-end security protocol on the network layer (Layer 3).
IPsec can be made transparent to end users since intermediate routers have no means of decrypting the packets.
Also, unlike others, the communication in IPsec usually uses multiple secure channels and has different keys to communicate with different destinations.
For these reasons, IPsec forms the basis of many Virtual Private Networking (VPN) solutions.
Services encrypted with IPsec are more independent from the infrastructure and flexibly deployable at many points of the network.

In IPsec, the entire IP packet is encrypted and authenticated by the initiator and a new IP header is added to route the packet to its destination.
The encrypted IP packets pass through the network without change until they reach the destination.
A major drawback of IPsec is its complexity and the introduced latency \cite{delay}.
It also significantly enlarges the size of the IP header, which causes network inefficiencies and adds penalties in the form of latency and reduced throughput to the overall solution cost.

\begin{figure}
  \includegraphics[width=\columnwidth]{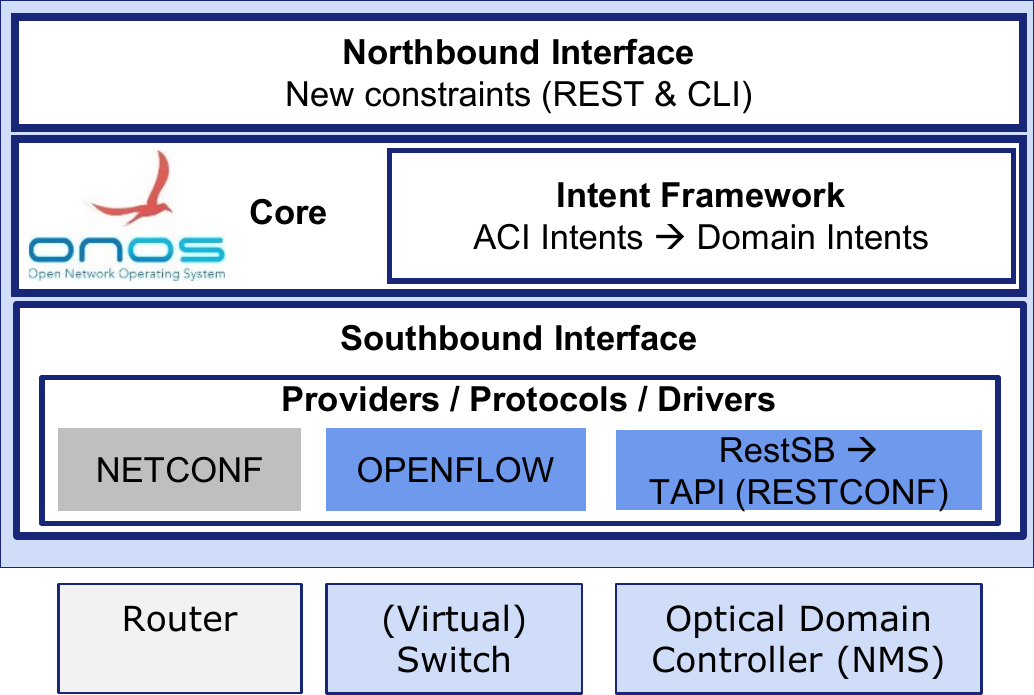}
  \caption{Basic system architecture of the ACINO orchestrator.}
  \label{fig:arch}
\end{figure}

\section{System Architecture}

\revision{The intent-based multilayer orchestrator, developed in the ACINO project and available at \url{https://github.com/ACINO-H2020}, is an open-source effort built on top of ONOS \cite{onos_hotsdn}.}
Many extensions, with the goal of making it more application centric, have been introduced to the original controller.
The ACINO orchestrator's simplified high-level architecture is presented in Fig.~\ref{fig:arch}.
Starting from the top, the intents, issued by a client application, are submitted through a REST northbound interface (NBI).
The orchestrator routes the request through the intent framework, compiles the submitted intent and selects the actions that need to be taken to satisfy the intent.
Those actions are mapped to the appropriate transport representation per device and sent through southbound protocols to the devices that need to be configured.
The devices themselves are either accessed directly or through a mediation layer like an intermediate controller, e.g., a network management system (NMS).
Protocols for southbound interactions include OpenFlow, NETCONF and RESTCONF.
The latter two define only the transport protocol and require YANG models for the description of the content, e.g., the ONF Transport API (TAPI) \cite{tapi}.
In any case, specialized drivers are implemented to handle the device specific behavior that is not covered by the common protocol definition.
The communication with the optical devices is done through ADVA’s NMS, which exposes an experimental TAPI interface.
The hardware side comprises optical equipment with physical layer encryption capabilities --- on a subset of the ports --- as well as switches which are able to install encrypted tunnels, i.e., MACsec.
Even though, routers could be easily included, they were omitted because of missing support for IPsec in the available hardware.

Next an example work flow for installing an intent is given to explain the process in more detail.
An application centric intent (ACI) is submitted through the NBI.
It defines requirements of the application, like bandwidth, latency and encryption.
This intent is handed over to a specialized compiler that is able to handle ACIs.
Since in the case of an optical network multiple devices are controlled through a single entry point, i.e., the mediation layer, the compiler needs to be able to create a special intent type, called domain intent.
Those domain intents are then installed through a protocol, e.g., RESTCONF, and the corresponding driver, e.g., TAPI.
Devices outside this domain, like switches, are still configured individually through OpenFlow.

\subsection{Domain Intents}

\begin{figure}[t!]
  \begin{lstlisting}[basicstyle=\small]{language=json}
  {
    "type": "AciIntent",
    "appId": "org.onosproject.cli",
    "priority": 100,
    "constraints": [
      {"type": "DomainConstraint"},
      {"type": "EncryptionConstraint"},
      {
        "type": "BandwidthConstraint",
        "bandwidth": 10000000
      }
    ],
    "one": "7E:1D:D7:77:7E:06/-1",
    "two": "CA:B8:53:D4:2A:84/-1"
  }
  \end{lstlisting}
  \caption{Request for an encrypted service in a JSON representation that can be submitted through the NBI.}
  \label{example}
\end{figure}

Domain intents are an important prerequisite for installing services in the optical domain.
\revision{They have been developed in the ACINO project in collaboration with other institutes, contributed to ONOS (\url{https://github.com/opennetworkinglab/onos}) and merged in version~1.10.0 (Kingfisher).}
The default approach in ONOS is to configure devices individually.
This method might be feasible in a network with switches and routers, but for optical networks and other domains managed by controllers it leads to increased complexity.
One reason is for example the setup of lightpaths through a network.
Besides configuring analog parameters, like launch power, the process itself needs a particular structure, e.g., equalization and hop-by-hop setup of lightpaths.
Additionally, a local controller might have a better knowledge about the domain because of additional information or a view without abstractions.
To handle intents for domains a specialized processing needs to be applied.
The main idea behind those intents is to configure parts of the network that are under control of a single controller and therefore, part of a domain.
A domain intent contains the ingress and egress points and might also contain information on a preferred path inside the domain.
It is assumed that the local controller is able to do all the required steps to fulfill the incoming requests.
Without an included path the domain controller can compute a valid path for provisioning.
In the second case, it needs to verify the feasibility of the requested path\,/\,configuration before installation.
In order to activate the domain processing an NBI intent needs to indicate this via a flag (\texttt{DomainConstraint} in Fig.~\ref{example}).

\subsection{TAPI Driver}

Even though ONOS provides numerous implementations of protocols out of the box, they only cover the common behavior in most cases.
To apply a protocol based on a YANG description, it is currently necessary to extend one of those protocols with the details of a particular implementation.
We implemented the Transport API (version 1.1) based on the definition that is available at github (\url{https://github.com/OpenNetworkingFoundation/Snowmass-ONFOpenTransport}).
The REST southbound protocol was used as a basis and starting point.
One advantage is that it supports intermediate controllers.
The TAPI is used for topology discovery, which includes nodes and links, as well as service setups.
The topology is exposed by the underlying controller, the individual elements are extracted and then exposed to ONOS by the driver.
Most information is directly mapped to available entities, some additional data is needed in order to be able to set up connections.
This information is stored in the form of annotations, compliant with ONOS' architectural requirements.

\begin{figure*}[!t]
  \centering
  \includegraphics[width=0.8\textwidth]{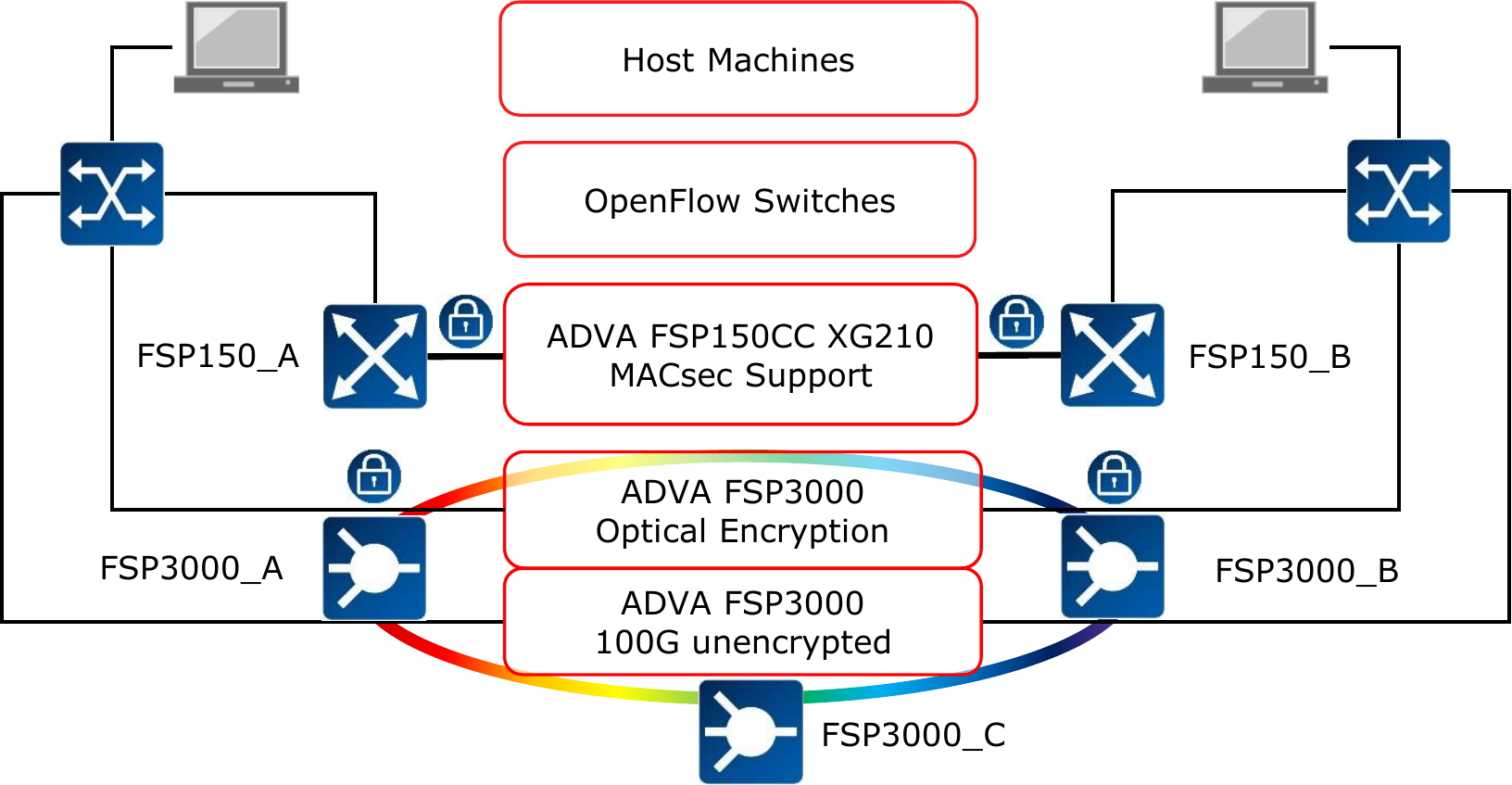}
  \caption{Testbed setup in the lab based on commercial networking hardware.}
  \label{fig:setup}
\end{figure*}

\subsection{Extensions for Encryption}

The task of introducing an intent-based secure service setup to ONOS, that automatically assigns the best layer of encryption, includes many steps.
Extensions start at the northbound interface with the definition of new primitives that allow the extended intent compiler to make a decision regarding the right layer of encryption for every request.
Then the compiler has to install the resulting secure service using new functionality in the drivers, including protocol extensions.
The enhancements of the existing ACINO orchestrator implementation \cite{netsoft} affected in particular the northbound interface, the intent processing and the southbound interface (SBI).
Like listed before, some of them have been already presented in \cite{ecoc_pdp,ofc_demo}.
Here, they will be explained in more detail.

At the NBI the changes revolved around the extension of the ACIs.
The encryption flag is a part of the intent that activates the intent processing that is needed for secure services.
Those intents can be submitted via ONOS' NBI or the command line interface (CLI).
One example for an application centric intent is shown in Fig.~\ref{example}.
It is noteworthy that the domain and encryption constraints are part of the request.
This initiates a special handling by the intent compiler.
The default components like source (``one''), destination (``two'') and bandwidth are reused for the encryption.

\revision{
  The ACI compiler was extended to support the processing of intents that want to establish a secure service.
  Considered constraints include availability, bandwidth and latency.
  A restriction or selection of the layer is considered a technological detail and therefore, not included in the intent.
  In the presented evaluation, the best layer for the encryption was chosen based on the bandwidth.
  The main reason is that we do not want to occupy more resources than needed, e.g., we use two optical ports and one path in total for a 10G connection instead of 20 1G Ethernet ports and 10 paths through the network.
  In addition, the two encryption mechanisms have similar properties in our testbed with respect to latency.
  In a multi-hop scenario this would be a differentiator because MACsec introduces latency by encrypting and decrypting at every hop along the path.
  Finally, we do not have any competing requests for the available resources in our evaluation, so that the availability is not an issue.
  A future implementation of the compiler needs to consider the limited resources and make a decision also based on availability.
  Therefore, secure services with a high bandwidth demand are mapped to physical encryption, while the ones with a lower demand are mapped to MACsec.
  If the encryption flag is missing, the (existing) unencrypted handling is applied.
  The compilation process also needs to take the properties of the available resources into account.
  For this reason the devices and ports have been annotated with information about encryption capabilities.
  This way the compiler can check if an option, satisfying all constraints, is available.
  If this is not the case, the intent is marked as failed and the user has to decide how to proceed.
  At the end, the compiler also has to communicate the need for an encryption to the underlying network.
}

The SBIs needed to implement new functionality to discover encryption capabilities and propagate encryption requests to the underlying hardware and mediation layer respectively.
For the retrieval of information about the encryption capabilities of the hardware, TAPI's label fields are used to indicate if a device\,/\,port is capable of providing encryption and on which layer.
For the service setup it is expected that the compiler chooses a correct pair of endpoints, otherwise the service setup and therefore, the intent will fail.

\section{Experimental Validation}

As shown in Fig.~\ref{fig:setup}, the presented system architecture was implemented and evaluated with commercial hardware in a lab.
The testbed comprised two off-the-shelf PCs for representing the hosts and two servers running the ADVA NMS and ACINO orchestrator.
In addition, the two PCs hosted an Open vSwitch (OVS) instance each.
Those virtual OpenFlow switches were used to steer the traffic into the right direction, depending on the encryption scheme, and were also under ONOS' control.
Both servers (not shown) were connected to the optical equipment as well as the OVS instances through a management network.
The testbed included an ADVA FSP3000 ROADM ring consisting of three nodes.
Two of them were equipped with 10G AES cards \cite{ADVA} which encrypt all traffic on the physical layer.
Also, two ADVA FSP150CC XG210 were part of the setup.
Those Ethernet demarcation devices are capable of encrypting the traffic using MACsec and were connected directly to each other.
The secret keys for the encrypted connections were preconfigured by the administrator.
For the unencrypted connection two 100G multiplexer cards were used.
We evaluated three scenarios, of which two requested a secure service and one needed an unencrypted connection.
For the first two, the best suited layer for the encryption was chosen automatically by the orchestrator.

\section{Measurements}

\begin{figure*}[!t]
  \centering
  \subfloat[Time from the submission of the intent until the compilation is finished.]{
    \includegraphics[width=0.31\textwidth]{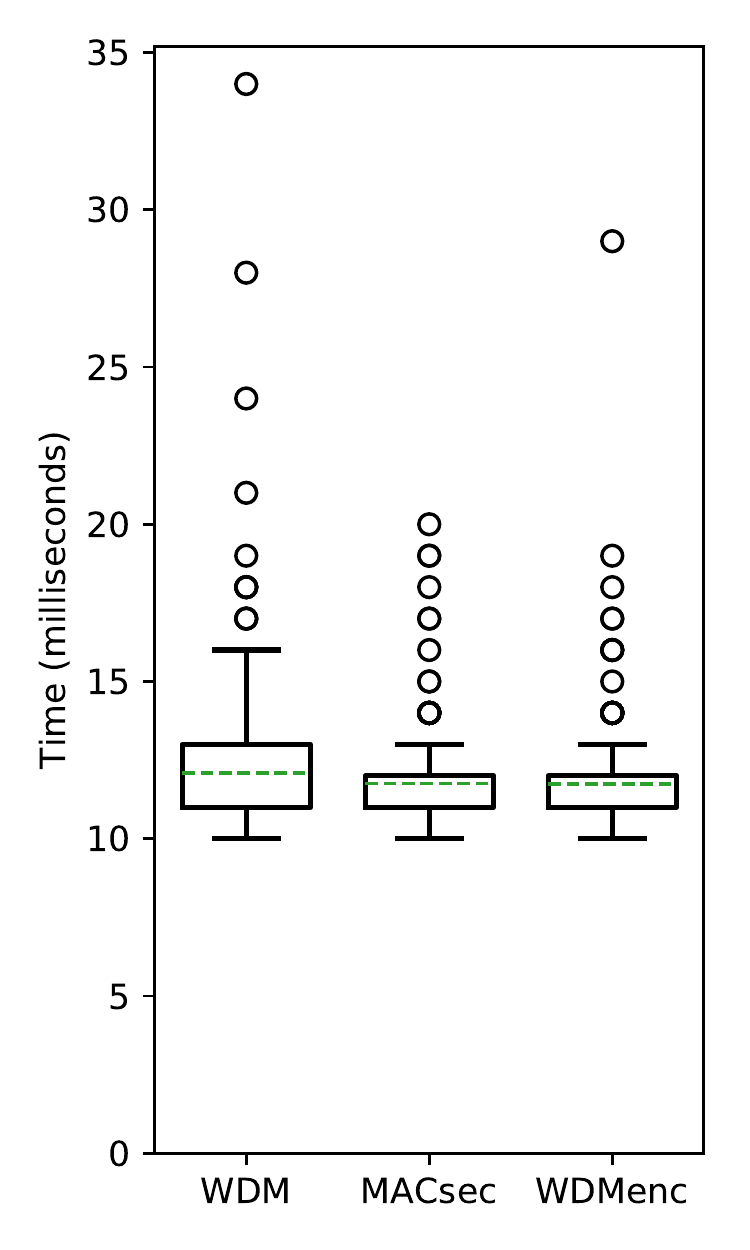}%
    \label{fig:measure:compile}
  }
  \hfil
  \subfloat[SBI installation time.]{
    \includegraphics[width=0.31\textwidth]{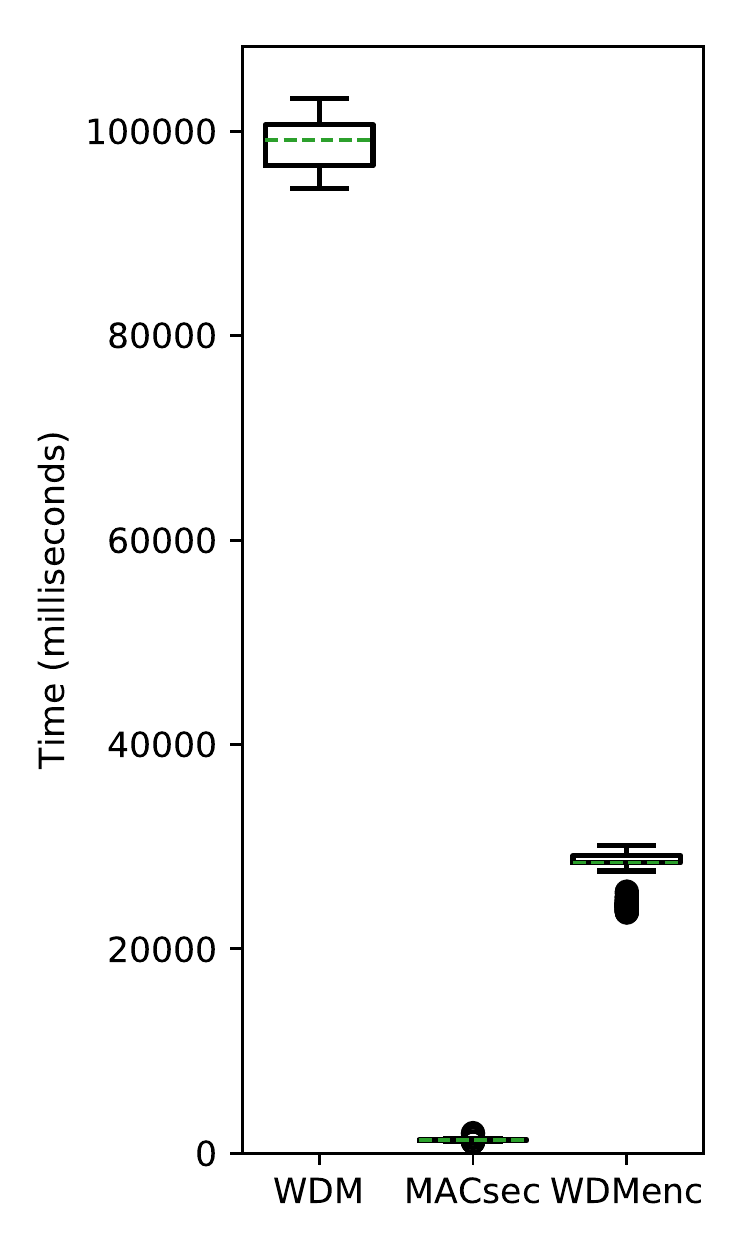}%
    \label{fig:measure:install}
  }
  \hfil
  \subfloat[SBI deletion time.]{
    \includegraphics[width=0.31\textwidth]{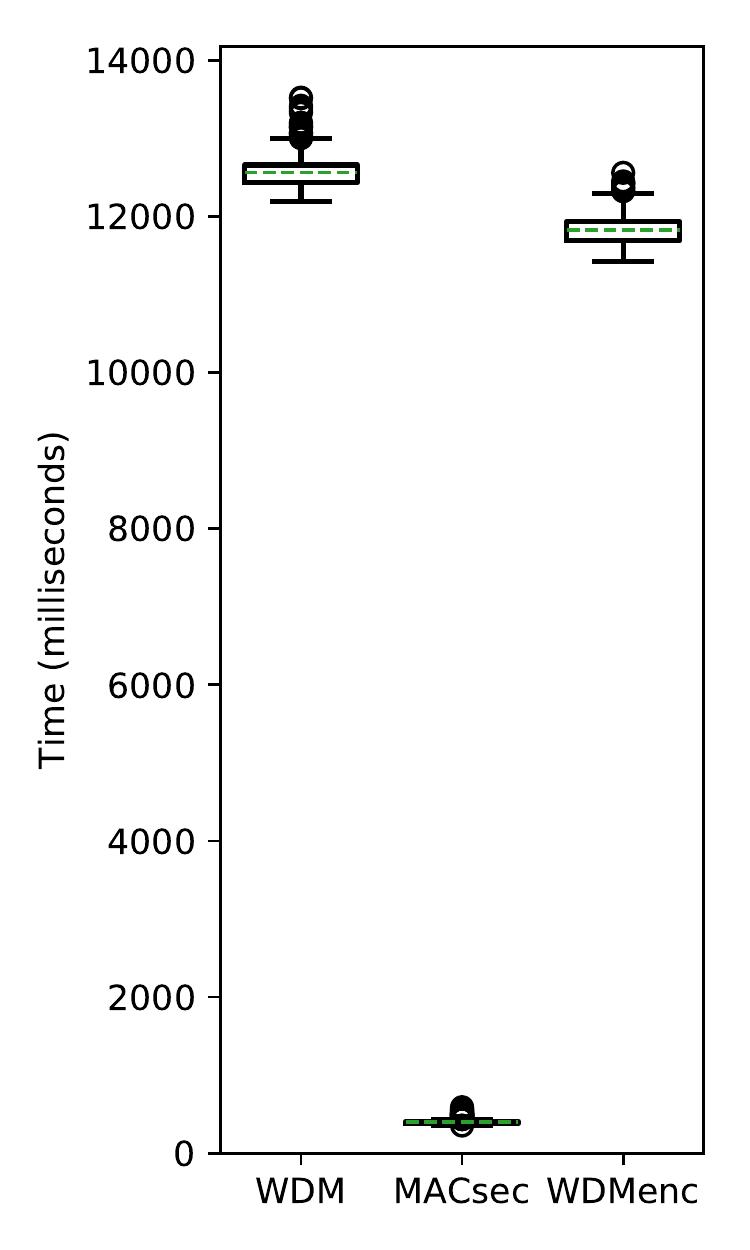}%
    \label{fig:measure:delete}
  }
  \caption{Measurements for the compilation, installation and deletion.}
  \label{fig:measure}
\end{figure*}

We conducted experiments for two layers of encryption, i.e., physical and MACsec, and for an unencrypted WDM connection.
Due to the software based implementation relying on OVS and therefore limited comparability, we omit the previously presented IPsec \cite{ecoc_pdp}.
We did more than 595 test runs for each of the three experiments.
The resulting graphs are presented in Fig.~\ref{fig:measure} and the results are discussed next.

First the time between the submission and the end of the compilation process was captured (see Fig.~\ref{fig:measure:compile}).
The goal was to evaluate if an additional delay was introduced by adding the encryption processing to the pipeline.
The measured times for the forwarding and processing of all three intents are about the same.
For the encrypted intents the mean values are 11.7~ms and for the unencrypted connection the mean value is 12.1~ms.
This might be related to the slightly different implementation of the unencrypted branch.
The margin of error for all of them is below 0.14~ms for a confidence level of 95\%.
The compilation time itself was below 1~ms on average and is only responsible for less than 10\% of the processing time.
As a result, we can state that the processing time remains about the same independent of the encryption handling.

The second graph (Fig.~\ref{fig:measure:install}) shows the time it took the SBI to complete the installation of the connection.
There is a big difference when it comes to the installation time of a connection.
For the unencrypted case, the mean value is about 99.1~s.
On the other hand, the encrypted WDM connection only needs 28.4~s.
\revision{
  The significant difference in setup time is a result of the used technologies.
  While the unencrypted multiplexed 100G connection uses coherent detection, the encrypted 10G applies direct detection.
  If both used the same technology, the results would be similar because the encryption has no influence on the setup time.
}
As expected, the MACsec setup, being based on an Ethernet connection, needs the least amount of time (1.3~s) for setting up the connection for a single hop.
The purpose of this graph is to visualize the difference in establishing a connection with growing capacity.
Switching from Ethernet to optical introduces a factor of more than 20.
Delays in setup might lead to a preference of another technology option and should be considered when fulfilling applications' intents.

The last graph (Fig.~\ref{fig:measure:delete}) captures the time it takes to delete an existing connection through the SBI.
Those values decide how long it takes before the resources are available again.
The teardown of a lightpath in the testbed takes 12.6~s for an unencrypted and 11.8~s for the encrypted connection.
Here, the difference between a 100G and a 10G connection is not very prominent.
With 401~ms, MACsec is also the fastest when it comes to the cleanup afterwards.
The conclusion is similar to the installation: the time for a deletion should be taken into account.
This time a factor of about 30 is introduced by switching to optical, although the absolute difference is lower than before.

In conclusion, even though the optical setup --- encrypted and unencrypted --- takes longer in the beginning, it is required for an efficient serving of higher bandwidths or to satisfy latency constraints.
Even from an economical point of view, it can be better to assign a dedicated lightpath instead of reusing higher layer transmission.
MACsec has some drawbacks on its own by applying encryption\,/\,decryption on every hop.
While the optical transmission can be used for long distances without intermediate processing, Layer 2 technologies need multiple hops to reach endpoints that are further apart.
Another result is that the type of appropriate encryption should also consider the setup and teardown times.
For short-lived connections or demands that need to be fulfilled as soon as possible optical connections might be unfit or need to be preprovisioned.

\section{Open Issues}

The demonstrated automatic creation of secure services presents a starting point for including data layer encryption in SDN.
Next we present a number of selected open issues that we came across during our work.

\subsection{Configurability vs. Ease of Use}
\revision{
  This work demonstrates how extensions to intents can be used so that a service is able to indicate its need for in-flight encryption.
  However, the definition of encryption for an application is not sufficient to specify requirements on mutual authentication mechanisms, key exchange protocols and algorithms for symmetric encryption.
  One possible solution is the use of application types in the intent specification, which are mapped to typical requirements in the respective field.
  A related idea is to provide (governmental) applications that rely on a high level of security with standard suites that are defined for example by the BSI (Germany) or the NIST (USA).
  The former proposal provides configurability, but can become cumbersome for many application groups, and the latter may not provide the level of detail required for specialized applications with strict security requirements.
  Further work is needed to evaluate the inherent trade-offs and define a basic vocabulary to address these scenarios.
}

\subsection{Operations, Administration, and Maintenance (OAM)}
\revision{
  Once provisioned, it is necessary to manage and monitor secure services.
  This paper demonstrates abstractions for encryption parameters that can be used for provisioning services.
  Similar abstractions need to be defined to address encryption specific OAM metrics in a layer agnostic fashion.
  The agnostic treatment of a service (and related OAM parameters) is out of scope of this work.
  We only propose an outline for metrics and events that can be used as a baseline to define such layer-agnostic encryption OAM parameters.
  Measurements like last successful\,/\,failed key exchange, error counters for failed key exchanges and time for which the service operates in transparent (unencrypted) mode can be used as generic measurement metrics.
  Encryption related events\,/\,alarms for successful\,/\,failed mutual authentication operations, key exchange sequences, configuration change notifications and internal (encryption module) failures can also be gathered independent of the layer.
  These parameters, coupled with operational state indicators and management operations can serve as an initial definition of OAM abstractions for layer-agnostic secure services.
}

\subsection{Security Issues}
\revision{
  The introduction of a service that is targeted toward secure applications is a likely candidate for malicious attacks.
  Security of controllers\,/\,orchestrators has been addressed in other studies \cite{sdn_sec_survey, controller_security} and is not part of this work.
  The security of the data plane poses fundamental challenges that need to be addressed in the future.
  One such example is the configuration of authentication mechanisms configured in the data plane, which are important to mutually confirm the identity of the communication partner.
  In this work, preconfigured keys in the controller and shared secret based authentication mechanisms were used at all encryption layers.
  The controller and the devices can mutually authenticate each other, but configuration of keys (shared secret key, public and private keys) from the application to the device should not include the controller as an intermediate entity.
  Mechanisms to do the same remain an open question in SDN.
  Another open challenge in the context of SDN and secure services, especially at layer 2 and below, is a mechanism to guarantee that the application traffic is using the configured secure path.
}

\subsection{Intent Negotiation}
\revision{
  In this work, the PoC implementation demonstrates that the orchestrator has an overview of all potential secure service mechanisms, and can provide a single solution to the application requesting the intent.
  The enforcement of constraints in this implementation is strict.
  The orchestrator will fail to offer a solution in case resources for encryption are not available in the selected layer chosen by the compiler.
  Application requirements, however, can be fuzzy.
  Intent negotiation mechanisms \cite{intent_demo, intent_ofc} can be used to interact with the user and identify a configuration which would be acceptable in case the originally requested secure service cannot be provisioned.
}

\section{Conclusion}
In future SDN controlled IP-over-optical networks, service requirements should be considered to provision services tailored to the application’s needs and to optimize network resource usage.
The presented extended ACINO orchestrator defines a lightweight northbound interface to specify the applications’ needs through intents.
In this paper we focus on security configuration, since the choice of an encryption concept depends on the intended needs.
We have shown an extension of our previously introduced automatic intent-based multilayer secure service creation.
The ACINO orchestrator creates multilayer secure services.
This approach has been experimentally verified and evaluated in a testbed with commercial optical equipment.
We have verified that, based on abstract requirements signaled by the application, the orchestrator can select service parameters in different layers and perform automatic multilayer secure service provisioning.
The performance measurements indicate that the additional time for processing security features is negligible.
In contrast, the setup and teardown times vary a lot for different technologies and hence should be considered when fulfilling applications' intents.
Since this approach is a generic concept, it could also be applied to future technologies, like quantum-secure encryption.


%



\section*{Acknowledgment}
This research has received funding from the European Commission within the H2020 Programme, ACINO project, Grant Number 645127.

\ifCLASSOPTIONcaptionsoff
  \newpage
\fi



\bibliographystyle{IEEEtran}
\bibliography{encryption}

\begin{thebibliography}{10}
\providecommand{\url}[1]{#1}
\csname url@samestyle\endcsname
\providecommand{\newblock}{\relax}
\providecommand{\bibinfo}[2]{#2}
\providecommand{\BIBentrySTDinterwordspacing}{\spaceskip=0pt\relax}
\providecommand{\BIBentryALTinterwordstretchfactor}{4}
\providecommand{\BIBentryALTinterwordspacing}{\spaceskip=\fontdimen2\font plus
\BIBentryALTinterwordstretchfactor\fontdimen3\font minus
  \fontdimen4\font\relax}
\providecommand{\BIBforeignlanguage}[2]{{%
\expandafter\ifx\csname l@#1\endcsname\relax
\typeout{** WARNING: IEEEtran.bst: No hyphenation pattern has been}%
\typeout{** loaded for the language `#1'. Using the pattern for}%
\typeout{** the default language instead.}%
\else
\language=\csname l@#1\endcsname
\fi
#2}}
\providecommand{\BIBdecl}{\relax}
\BIBdecl

\bibitem{livestats}
\BIBentryALTinterwordspacing
Internet live stats. Accessed on 01.09.2017. [Online]. Available:
  \url{http://www.internetlivestats.com/internet-users/}
\BIBentrySTDinterwordspacing

\bibitem{databreach}
\BIBentryALTinterwordspacing
IBM. Cost of data breach study. Accessed on 01.09.2017. [Online]. Available:
  \url{http://www-03.ibm.com/security/data-breach/}
\BIBentrySTDinterwordspacing

\bibitem{IPsec}
\BIBentryALTinterwordspacing
K.~Seo and S.~Kent, ``{Security Architecture for the Internet Protocol},'' RFC
  4301, Dec. 2005. [Online]. Available:
  \url{https://rfc-editor.org/rfc/rfc4301.txt}
\BIBentrySTDinterwordspacing

\bibitem{MACsec}
``Ieee standard for local and metropolitan area networks: Media access control
  (mac) security,'' \emph{IEEE Std 802.1AE-2006}, pp. 1--150, Aug. 2006.

\bibitem{ADVA}
\BIBentryALTinterwordspacing
{ADVA Optical Networking}. Fsp 3000 optical network encryption. Accessed on
  01.09.2017. [Online]. Available: \url{http://goo.gl/QrOn1W}
\BIBentrySTDinterwordspacing

\bibitem{archive}
\BIBentryALTinterwordspacing
Http archive. Accessed on 01.09.2017. [Online]. Available:
  \url{http://httparchive.org}
\BIBentrySTDinterwordspacing

\bibitem{EUCNC}
V.~Lopez, J.~M. Gran, J.~P. Fernandez-Palacios, D.~Siracusa, F.~Pederzolli,
  O.~Gerstel, Y.~Shikhmanter, J.~Mårtensson, P.~Sköldström, T.~Szyrkowiec,
  M.~Chamania, A.~Autenrieth, I.~Tomkos, and D.~Klonidis, ``The role of sdn in
  application centric ip and optical networks,'' in \emph{2016 European
  Conference on Networks and Communications (EuCNC)}, June 2016, pp. 138--142.

\bibitem{secaas}
V.~Varadharajan and U.~Tupakula, ``Security as a service model for cloud
  environment,'' \emph{IEEE Transactions on Network and Service Management},
  vol.~11, no.~1, pp. 60--75, March 2014.

\bibitem{sdn_sec_survey}
S.~Scott-Hayward, G.~O'Callaghan, and S.~Sezer, ``Sdn security: A survey,'' in
  \emph{2013 IEEE SDN for Future Networks and Services (SDN4FNS)}, Nov 2013,
  pp. 1--7.

\bibitem{cost_of_security}
R.~Durner and W.~Kellerer, ``The cost of security in the sdn control plane,''
  in \emph{ACM CoNEXT 2015 - Student Workshop}, Dec 2015.

\bibitem{controller_security}
S.~Scott-Hayward, ``Design and deployment of secure, robust, and resilient sdn
  controllers,'' in \emph{Proceedings of the 2015 1st IEEE Conference on
  Network Softwarization (NetSoft)}, April 2015, pp. 1--5.

\bibitem{sdn_threats}
D.~Kreutz, F.~M. Ramos, and P.~Verissimo, ``Towards secure and dependable
  software-defined networks,'' in \emph{Proceedings of the Second ACM SIGCOMM
  Workshop on Hot Topics in Software Defined Networking}, ser. HotSDN
  '13.\hskip 1em plus 0.5em minus 0.4em\relax New York, NY, USA: ACM, 2013, pp.
  55--60.

\bibitem{eavesdropping}
J.~Spooner and S.~Y. Zhu, ``A review of solutions for sdn-exclusive security
  issues,'' \emph{International Journal of Advanced Computer Science and
  Applications (ijacsa)}, vol.~7, no.~8, 2016.

\bibitem{optical_security}
M.~Furdek, N.~Skorin-Kapov, S.~Zsigmond, and L.~Wosinska, ``Vulnerabilities and
  security issues in optical networks,'' in \emph{2014 16th International
  Conference on Transparent Optical Networks (ICTON)}, July 2014, pp. 1--4.

\bibitem{ecoc_pdp}
T.~Szyrkowiec, M.~Santuari, M.~Chamania, D.~Siracusa, A.~Autenrieth, and
  V.~Lopez, ``First demonstration of an automatic multilayer intent-based
  secure service creation by an open source sdn orchestrator,'' in \emph{42nd
  European Conference on Optical Communication (ECOC2016)}, September 2016, pp.
  1--3.

\bibitem{ofc_demo}
M.~Chamania, T.~Szyrkowiec, M.~Santuari, D.~Siracusa, A.~Autenrieth, V.~Lopez,
  P.~Sköldström, and S.~Junique, ``Intent-based in-flight service encryption
  in multi-layer transport networks,'' in \emph{2017 Optical Fiber
  Communications Conference and Exhibition (OFC)}, March 2017, pp. 1--2.

\bibitem{MACsec:auth}
``Ieee standard for local and metropolitan area networks--port-based network
  access control,'' \emph{IEEE Std 802.1X-2010 (Revision of IEEE Std
  802.1X-2004)}, pp. 1--205, Feb. 2010.

\bibitem{delay}
R.~Ramaswamy, N.~Weng, and T.~Wolf, ``Characterizing network processing
  delay,'' in \emph{Global Telecommunications Conference, 2004. GLOBECOM '04.
  IEEE}, vol.~3, Nov 2004, pp. 1629--1634 Vol.3.

\bibitem{onos_hotsdn}
P.~Berde, M.~Gerola, J.~Hart, Y.~Higuchi, M.~Kobayashi, T.~Koide, B.~Lantz,
  B.~O'Connor, P.~Radoslavov, W.~Snow, and G.~Parulkar, ``Onos: Towards an
  open, distributed sdn os,'' in \emph{Proceedings of the Third Workshop on Hot
  Topics in Software Defined Networking}, ser. HotSDN '14.\hskip 1em plus 0.5em
  minus 0.4em\relax New York, NY, USA: ACM, 2014, pp. 1--6.

\bibitem{tapi}
\BIBentryALTinterwordspacing
C.~Qiaogang, E.~Segev, E.~Varma, G.~Zhang, H.~Ding, I.~Busi, J.~He,
  K.~Sethuraman, L.~Ong, N.~Davis, R.~Vilalta, S.~Bellotti, and V.~Lopez,
  ``{Functional Requirements for Transport API},'' Open Networking Foundation,
  Technical Recommendations ONF TR-527 (Version No.01), Jun. 2016, work in
  Progress. [Online]. Available:
  \url{https://www.opennetworking.org/images/stories/downloads/sdn-resources/technical-reports/TR-527_TAPI_Functional_Requirements.pdf}
\BIBentrySTDinterwordspacing

\bibitem{netsoft}
M.~Santuari, T.~Szyrkowiec, M.~Chamania, R.~Doriguzzi-Corin, V.~Lopez, and
  D.~Siracusa, ``Policy-based restoration in ip/optical transport networks,''
  in \emph{2016 IEEE NetSoft Conference and Workshops (NetSoft)}, June 2016,
  pp. 357--358.

\bibitem{intent_demo}
A.~Marsico, M.~Santuari, M.~Savi, D.~Siracusa, A.~Ghafoor, S.~Junique, and
  P.~Skoldstrom, ``An interactive intent-based negotiation scheme for
  application-centric networks,'' in \emph{2017 IEEE Conference on Network
  Softwarization (NetSoft)}, Jul. 2017, pp. 1--2.

\bibitem{intent_ofc}
A.~Marsico, M.~Savi, D.~Siracusa, and E.~Salvadori, ``An automated
  service-downgrade negotiation scheme for application-centric networks,'' in
  \emph{2018 Optical Fiber Communications Conference and Exhibition (OFC)},
  March 2018, pp. 1--3, to be published.

\end{thebibliography}
\end{document}